\documentclass[10pt, conference, letterpaper]{IEEEtran}
\usepackage{algorithm}
\usepackage{algpseudocode}
\makeatletter
\newcommand{\algmargin}{\the\ALG@thistlm}
\makeatother
\newlength{\whilewidth}
\algdef{SE}[parWHILE]{parWhile}{EndparWhile}[1]
  {\parbox[t]{\dimexpr\linewidth-\algmargin}{%
     \hangindent\whilewidth\strut\algorithmicwhile\ #1\ \algorithmicdo\strut}}{\algorithmicend\ \algorithmicwhile}%
\algnewcommand{\parState}[1]{\State%
  \parbox[t]{\dimexpr\linewidth-\algmargin}{\strut #1\strut}}
\usepackage{lipsum}
\usepackage{amsmath}
\usepackage{amsfonts}
\usepackage{amssymb}

\usepackage[]{graphicx} 
\usepackage{cite}
\usepackage{amsthm}
\usepackage{epstopdf,epsfig}
\usepackage{xcolor}
\usepackage{balance}
\usepackage{colortbl}
\usepackage{subcaption}
\usepackage{tikz}
\usetikzlibrary{plotmarks}
\usepackage{pgfplots}
\usepackage{mathrsfs}
\newcommand{\MU}[1]{MU~$#1$}
\newcommand{\Ksize}{K}
\newcommand{\scr}[1]{\mathscr{#1}}
\newtheorem{lemma}{Lemma}

\newtheorem{definition}{Definition}

\IEEEoverridecommandlockouts
\begin{document}
\title{Energy-Aware Optimization and Mechanism Design for Cellular Device-to-Device Local Area Networks}
\author{Mehdi~Naderi~Soorki$^{1,2}$,
	Mohammad Yaghini$^{1}$,
        Mohammad Hossein~Manshaei$^{1}$,
	Walid~Saad$^{2}$,
       and~Hossein~Saidi$^{1}$\\
        \small $^{1}$ Department of Electrical and Computer Engineering, Isfahan University of Technology, Isfahan, 84156-83111, Iran, \\
         Email: \{m.naderisoorki,m.yaghini\}@ec.iut.ac.ir,\{manshaei,hsaidi\}@cc.iut.ac.ir \\
       \small $^{2}$ Department of Electrical and Computer Engineering, Virginia Tech, Blacksburg, VA, USA, Email: \{mehdin,walids\}@vt.edu
      \thanks{This research was supported by the U.S. National Science Foundation under Grant CNS-1513697.}
       }

\maketitle

\begin{abstract}
In a device-to-device (D2D) local area network (LAN), mobile users (MUs) must cooperate to download common real-time content from a wireless cellular network.
However, sustaining such D2D LANs over cellular networks requires the introduction of mechanisms that will incentivize the MUs to cooperate.
In this paper, the problem of energy-aware D2D LAN formation over cellular networks is studied. The problem is formulated using a game-theoretic framework in which each MU seeks to minimize its energy consumption while actively participating in the D2D LAN.
To account for the selfish behavior of the MUs, a punishment and incentive protocol is proposed in order to ensure cooperation among MUs. Within this protocol, an estimation algorithm is proposed to simulate the process of D2D LAN formation and, then, adjust the mechanism parameters to maintain cooperation. Simulation results show that the proposed framework can improve energy efficiency up to 36\% relative to the traditional multicast scenario.
\end{abstract}

\section{Introduction}
\label{sec:Intro}
D2D communications over the licensed cellular spectrum is viewed as an important feature that will enable emerging 5G cellular systems to deliver high-speed data rates~\cite{Asadi2014}. D2D communication enables direct communications between mobile users (MUs) without going through the base stations (BSs).

A class of D2D communication is device-to-device local area network (D2D LAN). In multi-hop D2D LAN, network-controlled smart devices can realize cluster-based communication in an ad hoc manner while operating on the cellular band. Using D2D LAN, data requests can be offloaded efficiently. 
In D2D LAN, MUs actively transmit and receive data on two wireless links: one link to communicate with the BS over the cellular band, known as the long range (LR), and another link to communicate with other MUs in the D2D LAN, known as the short range (SR) link.

Group communication is an application scenario of D2D LAN in which the BS receives a large number of requests for similar data.
One class of this emerging mobile application is simultaneous multicasting of common real-time content to a group of MUs over the cellular band~\cite{alexact}.

Conventionally, when MUs request the same content, the BS can multicast it to all MUs. But in a D2D LAN, the BS only multicasts the content to a group of MUs, known as ``the seeds''~\cite{alexact}, who in turn forward it to the rest of MUs.

Several studies have proposed cellular offloading via D2D communication and the main goal of existing literature is reducing the energy consumption of MUs~\cite{alexact,al2014optimal} and~\cite{al2008optimal}. The problem of finding the optimal energy-aware one-hop or multi-hop cooperation for all unicast/multicast combinations on LR and SR links is  NP-hard as shown in~\cite{al2014optimal}.  Regarding the cooperation model, existing studies assume full cooperation, centralized location-based group formation, or distributed group formation\cite{al2008optimal,popova2008cooperative}.

In~\cite{al2014optimal}, authors attempt to optimize the MUs' total energy consumption by selecting optimal seeds. However, they assume that MUs are ``altruistic'' and therefore such works do not study the incentives for participation in a D2D LAN.

The main contribution of this paper is to propose a novel energy-aware protocol, dubbed \emph{multi-hop collaborative real-time content dissemination} (MCRCD),  for enabling multi-hop D2D LAN formation among selfish MUs.
We formulate the problem as a graph formation problem while considering the MUs' rational and selfish behavior using a game-theoretic model. We also propose incentive and punishment mechanisms for maintaining cooperation among MUs for  content distribution. Moreover, in order to adjust the parameters of the mechanisms, we propose an estimation algorithm. We show that, using MCRCD, forming a sustainable D2D LAN is possible. Simulation results in significant improvements in the energy efficiency.
In addition to improving energy-efficiency, we look for an efficient scheduling scheme which selects seeds in a way that encourages cooperation among MUs.

The rest of this paper is organized as follows. In Section \ref{Sec:SystemModel}, we present the system model. The MCRCD protocol is discussed in Section \ref{Sec:MCRCD}.
Simulation results are presented in ‎Section \ref{Sec:Simulation} and conclusions are drawn in Section \ref{Sec:Conclusion}.
\section{System Model}
\label{Sec:SystemModel}
Consider the downlink of an LTE network in which OFDMA is used to share the spectrum between various users, using LR links. In this system, MUs can also communicate with one another using SR links. The system bandwidth $B$ is divided into $X$ resource blocks (RBs); each of which contains $\alpha$ subcarriers. We consider an infinite number of discrete time slots $\{1, 2,\dotsc, t,\dotsc\}$; each of duration $T$. In this system, a number of D2D LANs can communicate over the cellular spectrum.
Let $\mathcal{K}_t$ be the set of all MUs present at time slot $t$. A D2D LAN formed between a subset of MUs will be represented by a directed graph $G^t_i$ whose vertices are the MUs and whose edges are the connections between them.
Within each D2D LAN $G_i^t$, one specific MU $i$ is selected to be a \emph{seed} which receives the data directly from the BS via an LR link and, then, it multicasts this data over its SR link.
In this model, we assume that every MU is selected to be the seed for a period $\rho_iT$ during each time slot.
An MU can join or leave a D2D LAN only at the beginning or the end of each time slot. Upon reception of the real-time content on the LR link, the seed multicasts the data, using its SR link, to its neighboring MUs (indicated by the graph) in a single-hop manner. Subsequently, these one-hop neighbors multicast the data to their neighbors located at a two-hop distance from the seed. This process continues for a maximum of $H$ hops, until all members of the D2D LAN receive the data.

At any given time, an MU can have one of three roles: seed, relay or sink. A relay receives data on its SR link and multicasts it also on its SR link while a sink only receives on its SR link and does not relay. Non-selfish MUs can form the optimum D2D LAN $\widehat{G_{i}}$ that minimizes MUs' total energy consumption~\cite{al2014optimal}. However, in such a D2D LAN, selfish MUs are not motivated to relay and as a result, the content must be multicast to every MU with a rate limited by the worst channel condition between the MUs and the BS.

An alternative is to provide incentives for selfish MUs to form a D2D LAN.
At each time slot $t$, every \MU{i} is assigned a specific time duration $\rho_i T$ with $\sum\rho_i = 1$, referred to as the ``seed time'', in which MU $i$ will be the seed of the D2D LAN. As discussed in Subsection \ref{sec:optimization}, $\rho_i$ can be set in a manner that provides incentive for participation in the D2D LAN.
Following the selection of an MU as the seed, others will self-organize into a graph without any assistance from the BS.
The seed and the relays then decide whether to cooperate (relay) or defect (refrain from relaying).

\begin{figure}[!t]
\centering
\includegraphics[width=0.5\textwidth]{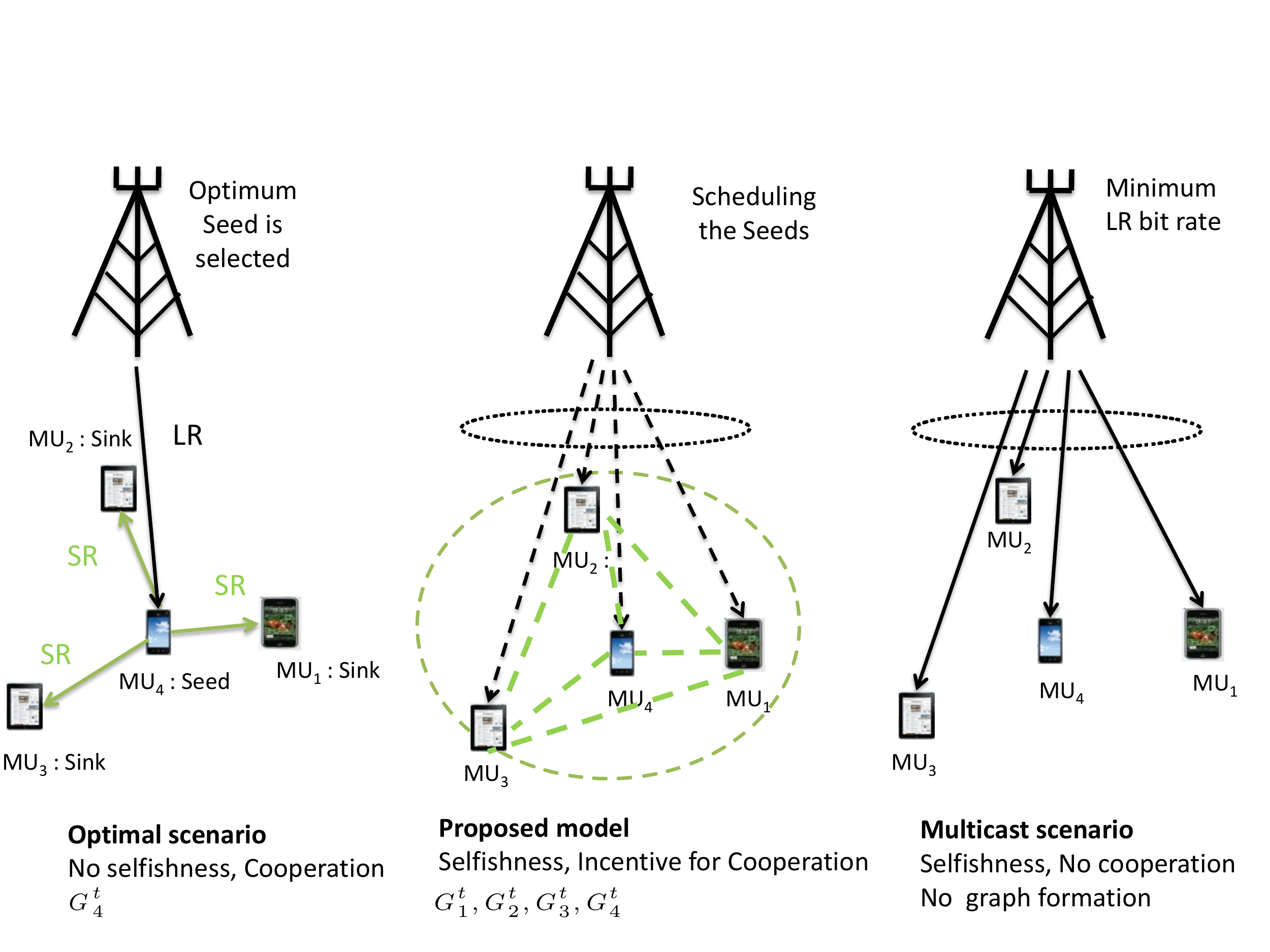}‎
\caption{An illustration of optimal scenario, MCRCD protocol and multicast scenario of \MU{1}, \MU{2}, \MU{3} and \MU{4}.}
  \vspace{-1.5em}
\label{system model}
\end{figure}

Fig. \ref{system model} illustrates the optimal, multicast, and the proposed scenarios for receiving common real-time content. In Fig. \ref{system model}, at time slot $t$, all MUs are within the range of one another.
In the optimal scenario, where MUs' total energy consumption is minimized and they are not selfish, a D2D LAN between MUs $1$, $2$, $3$, and $4$ will form. \MU{4}~is selected as the optimal seed for the entire duration of $T$ and the optimal graph $\widehat{G_{4}^t}$ is formed; in which, \MU{1}, \MU{2} and \MU{3} are sinks.
However, if MUs are selfish and non-cooperative the multicast scenario takes place and all MUs receive on their LR link with the minimum bit rate.
In the proposed scenario, the BS sequentially schedules MUs to be the seed for specific fractions of $T$ ($\rho_iT$) where $\sum_{i=1\dotsc4}\rho_iT=T$.
The resulting graphs $G_i^t,\;i=1\dotsc4$, are illustrated in Fig. \ref{SModel2}.

{\textbf{Notation:} }{In what follows, D2D LAN parameters have primes,~$E'$ and $X'$, while the same parameters for multi-casting use regular letters.
Moreover, energy, power and rate parameters are in calligraphic letters for SR links ($\mathscr{E}$, $\mathscr{P}$ and $\mathscr{R}$) and in regular letters for LR links ($E$, $P$ and $R$).

\begin{figure}[t]
\centering
\includegraphics[height=1.8in, width=0.5\textwidth]{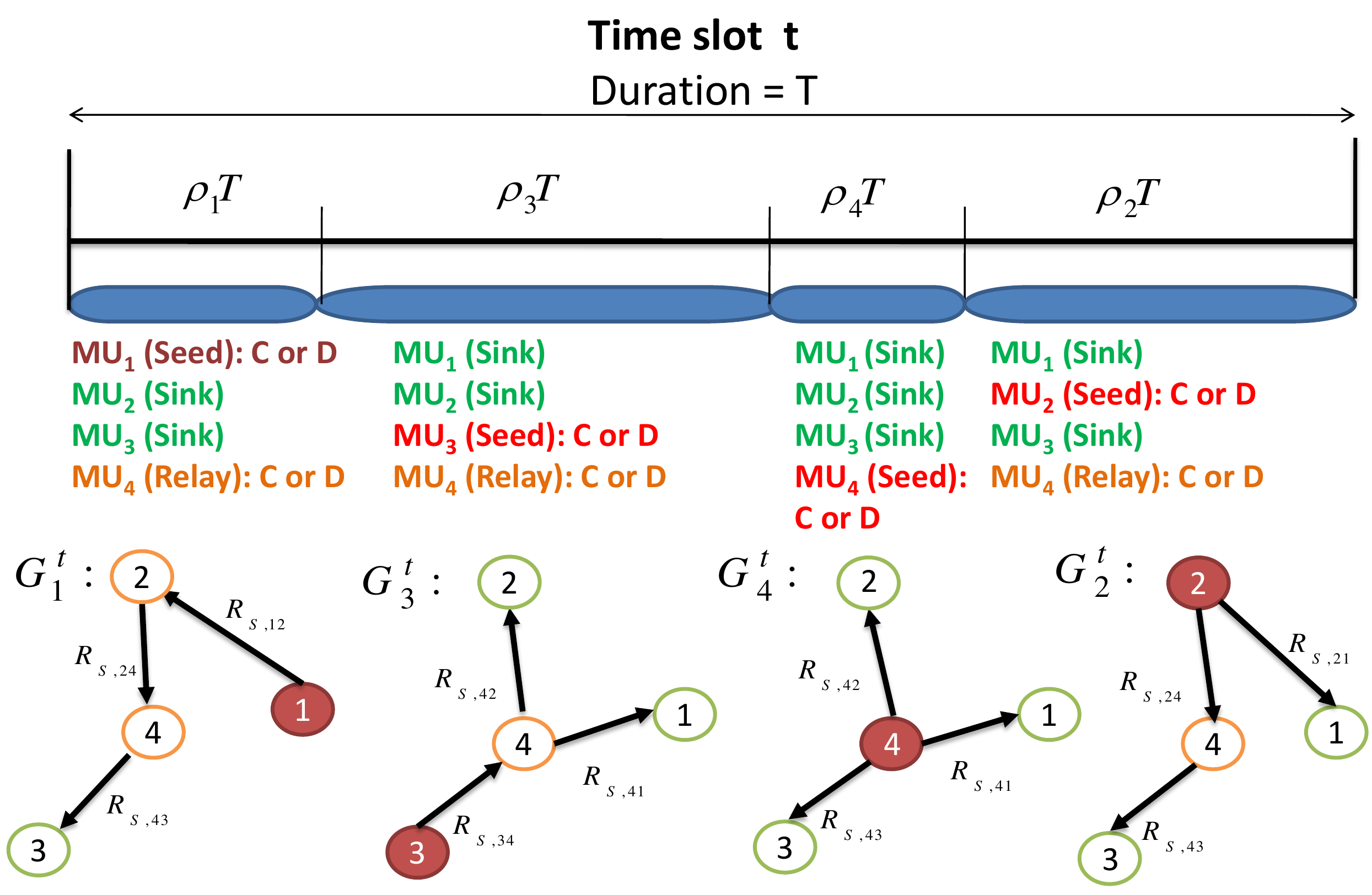}
\caption{Every MU is selected to be the seed during a portion of the time slot. With every seed, a new graph will form.}
\label{SModel2}
\vspace{-1.5em}
\end{figure}
\subsubsection{Bit Rate}
\label{sec:bit_rate}

The bit rate between an \MU{k} and the BS is given by~\cite{goldsmith2005wireless}: $R_{k} =\sum_{z\in{\mathcal{X}_k}} B_z \log_2 (1+\frac{\beta P^{z}h_{k}^z}{\sigma^2+\sum_{i\neq k}{P_i^z h_{ik}}})$, where $B_z$ is the bandwidth of each subcarrier which equals $\frac {B}{\alpha X}$, and $\mathcal{X}_k$ is the set of subcarriers assigned to \MU{k} on the cellular LR link, and $\beta=\frac{1.5}{\ln(5P_e)}$ is the SNR gap for M-QAM modulation with $P_e$ being the maximum acceptable error probability~\cite{goldsmith2005wireless}. The $P^z$ is the BS transmitter power over subcarrier z and $h_k^z$ is the channel gain between \MU{k} and the~BS.
If the BS conventionally multicasts common data to all requesting MUs, the bit rate on the LR link will be limited to the worst channel condition between the MUs and the BS. Therefore, we have $R=\min_{k\in \mathcal{K}_t} R_{k},$ where $\mathcal{K}_t$ is the set of all MUs that request a common data at time slot $t$ and don't form a D2D LAN.
We assume an underlaid D2D scheme that uses device relaying with operator controlled link establishment for device-tier communications~\cite{Asadi2014}. The BS handles resource allocation (channel assignment over the cellular band and power allocation for the SR links in the D2D tier). The bit rate between two MUs $k$ (transmitter) and $j$ (receiver) on the D2D link can be calculated as follows~\cite{goldsmith2005wireless}: $\scr{R}_{k,j} =\sum_{z\in X'} {B_z}\log_2 {(1+\frac{ \scr{P}_k^z h_{kj}^z}{ I+\sigma^2})}$, where $X'$ is the set of RBs that the BS assigns to the D2D LAN.
$I$ is the interference that \MU{k} experiences on each sub-carrier z in the D2D LAN and is equal to
$\sum_{i\neq k}{\scr{P}_i^z h_{ij}}+{P^z h_{j}}$.
The first term indicates the interference from other D2D pairs and the second term is due to the interference from the BS. The latter is only present if the BS uses the same sub-channels for both SR D2D communications and LR cellular communications, which is the case for underlaid D2D communication~\cite{Asadi2014}.
When \MU{m} is selected as the seed, the multicast bit rate of \MU{k} on the SR link will be given by $\scr{R}_{k} =\min_{j \in X, kj\in G_m} \scr{R}_{kj}$.
Thus, at time slot $t$, $\scr{R}_k$ depends on the graph $G_m$ and is restricted by the worst bit rate between \MU{k} and its one-hop neighbors
in $G_m$. The reception bit rate of \MU{k} is $\scr{R}_{k}=\scr{R}_{s_k}$,
in which \MU{s_k} is the \textit{source} of \MU{k} in $G_m$, i.e., the node that transmits the content to \MU{k}. Note that the graph of a D2D LAN is a tree, i.e., a graph without simple cycles.
\subsubsection{Energy Consumption}
The energy consumption generally depends on transmission/reception time
and bit rate~\cite{albano2011context}. For our network, we apply the generic energy consumption model of~\cite{al2014optimal} in which, the energy consumption is formulated as $P(R)T$. $P(R)$ is the energy consumed per second while transmitting or receiving with bit rate $R$ and $T$ is the time duration in seconds. We do not consider a power allocation mechanism and assume all MUs are transmitting at their maximum power $P_t$ and that the base station transmits with constant power $P_{Tx,L}$ on all of its subcarriers.
When MUs do not form a D2D LAN, during the entire time slot, every MU should download the content using its LR link.
Therefore the energy consumption of \MU{k} will be:
\begin{equation}
E_k=P_{Rx}T,
\label{Emulti}
\end{equation}
where $P_{Rx}$ depends on $R_{Rx}$. Since the role of each MU (seed, relay or sink) is subject to change during each time slot, to determine the total energy consumption of an MU, we need to calculate the power consumed when an MU takes on these roles. In a D2D LAN, the role of each MU depends on which MU is selected as the seed as well as the graph that is formed. In the following we assume that \MU{m} is selected as the seed, therefore graph $G_m$ is formed during $\rho_m T$. The energy consumption of the seed includes the energy to receive data on the LR link as well as the energy to multicast the data
to its one-hop neighbors. In this case $m=k$ thus,
\begin{equation}
E_{k}(G_k)=(P_{Rx}+\scr{P}_{Tx}) \rho_k T,
\label{Eseed}
\end{equation}
in which $P_{Rx}$ and $\scr{P}_{Tx}$ depend on $R_{Rx}$ and $\scr{R}_{Tx}$, respectively. The energy consumption of the relay is the sum of the energy spent to receive the data from the source (the seed or another MU)
using the SR link, and the energy spent to multicast it to other MUs using the same link. MU $m\neq k$ is selected as the seed and \MU{k} is a relay in $G_m$. Therefore,
\begin{equation}
E_{k}(G_k)=(\scr{P}_{Rx}+\scr{P}_{Tx}) \rho_m T,
\label{Erelay}
\end{equation}
in which, $\scr{P}_{Rx}$ and $\scr{P}_{Tx}$ depend on $\scr{R}_{k}$.
A sink only receives data on its SR link.
If \MU{m}, $m\neq k$ is the seed and \MU{k} is a sink, it is true that $\forall j \in G_m \text{, }j\neq k$, $G_m$ does not have the edge $(k,j)$. Therefore,
\begin{equation}
E_{k}(G_m)=\scr{P}_{Rx} \rho_m T,
\label{Esink}
\end{equation}
in which $\scr{P}_{Rx}$ depends on $\scr{R}_{Rx,k}$. The role of \MU{k} changes according to the formed graph.
The total energy consumption of \MU{k} during time slot $t$ is
\begin{equation}
E_{k}= \sum_{j=1}^{\Ksize}E_k(G_j),
\label{Etotal}
\end{equation}
where $E_k(G_j)$ is given by (\ref{Eseed}), (\ref{Erelay}), or (\ref{Esink}) based on its role following the selection of the seed at time slot $t$.

\section{The MCRCD Protocol}
\label{Sec:MCRCD}
Maintaining cooperative behavior is a challenge in D2D LANs due to the fact that MUs often have to consume energy to assist others in the LAN, especially for seed or relay MUs. One approach to design such incentives is via the use of non-cooperative game theory \cite{shoham2008multiagent}. In our model, we assume that the selfish MUs in  $\mathcal{K}_t$ trust the BS as a trusted third party (TTP) and will abide by its decisions. Thus, the role of the BS is that of a mediator in the game between MUs which is suitable when dealing with D2D over cellular.
The MCRCD protocol has four phases: Distributed neighbor discovery, graph estimation and scheduling mechanism, self-organizing D2D LAN formation and self-punishment mechanism.
These phases are sequentially repeated during each time slot in order to track changes in the environment (due to the mobility of MUs) or prior self-punishment phases.

During the first phase, distributed neighbor discovery, the MUs autonomously perform a distributed neighbor discovery protocol and generate $\mathcal{K}_t$. Having found their neighbors, each MU can estimate the channel conditions with its neighbors using methods such as those discussed in~\cite{kim2014full}. Next, MUs will truthfully feedback this information to the BS. The remaining phases are as follow.

\subsection{Graph Estimation and Scheduling}
\label{sec:GE}
The second phase is performed by the BS. Before we start discussing this phase, we need to introduce the process of graph formation which, in effect, belongs to Phase III and is carried out by the MUs and not the BS. This is necessary because in this phase, the BS essentially tries to estimate the outcome of Phase III in order to tweak the parameters of the scheduling mechanism later on. When MUs start to form the D2D LAN, each wants to connect to a peer with whom it would have the highest possible bit rate. This is because as the bit rate $R$ increases, the energy per bit $E(R)/R$ decreases\cite{albano2011context}. Therefore, each \MU{i} maintains a preference vector of other MUs that it sorts based on the bit rates between itself and other MUs. In order to receive the content using the least amount of energy, \MU{i} proposes to connect to its favorite peer, \MU{j}, who accepts the connection only if the connection can be maintained in real time with no buffering due to low bit rate on the receiving~side. In order to properly adjust the mechanism parameters, namely $\rho_i$ and the seed \MU{k}(t),  the BS must have an estimation of the behavior of MUs who interact using the aforementioned scheme.
Thus, we propose a heuristic graph~estimation algorithm that effectively mimics the behavior of~MUs.
\subsubsection{Graph Estimation Algorithm}
Algorithm \ref{alg:estimation} is a centralized algorithm that is used to estimate the outcome of D2D LAN formation, given that some \MU{k} is chosen as the seed.

In the MCRCD protocol, the BS assumes that, after selecting the seed, MUs autonomously form the graph of D2D LAN based on the transmission bit rates between themselves on the SR links, as we outlined in the beginning of this subsection. To track the process of proposals and acceptances, the BS maintains several sets and vectors: the accepted set $\mathcal{A}$, the Unconnected set $\mathcal{B}$, the $1\times K-1$ proposal order vector~$\boldsymbol{p}$ and the $K-1\times K-1$ matrix~$\boldsymbol{P}$ that represents MUs' preferences.

In the beginning, the seed is the only member of $\mathcal{A}$ and all other MUs are in $\mathcal{B}$. Therefore, when their turn arrives (signaled by $\boldsymbol{p}$) MUs can ``propose''  only to the seed.
Since the content is in real time, if the proposing MU's connection is weak, the seed would have to compensate for this by buffering the content. This is not acceptable, since seeds are also mobile units with limited memory and battery. Consequently, the seed will only accept a connection from \MU{i} if the condition $\scr{R}_{ki}\geq R_{k}$ is satisfied. For the rest of the process, $\mathcal{A}$ contains more than one MU, namely, the seed and some relays. Thus, the remaining MUs in $\mathcal{B}$ propose to the current members of $\mathcal{A}$. In this case, an \MU{j} will only accept the connection to \MU{i} if a similar  condition is satisfied: $\scr{R}_{ji}\geq \scr{R}_{js_j}$,
where $s_j$ is the source of \MU{j}, i.e., the MU that feeds \MU{j} itself. This process of proposing and accepting/rejecting continues until every MU has a path to the seed. Due to sequential nature of this process, the order of proposals affects the outcome of the graph formation; thus the need for $\boldsymbol{p}$. This vector is announced by the BS at the beginning of each time slot and indicates when each MU must propose.
$\boldsymbol{p}$ can then be permutated for the next time slot to ensure fairness. We note that $\boldsymbol{p}$ can potentially be used as yet another incentive/punishment mechanism. In summary, \MU{i} proposes to its favorite MU based on the $i$th row of $\boldsymbol{P}$ (which represents its sorted preference list) and all such MUs should wait their turn indicated by $\boldsymbol{p}$.
{\scriptsize
\begin{algorithm}[t]
\begin{algorithmic}
\caption{ Graph Estimation Algorithm‎}
\label{alg:estimation}
\State{\bfseries input. }$\mathcal{K}_t$, \MU{k} (the seed), $\boldsymbol{p}$, $\boldsymbol{P}$
\State{\bfseries output. } The formed graph \emph{$G_k$}
\vspace*{1mm}
\State{{\bfseries Step 1.} Initialize  $\mathcal{A}= \{\text{\MU{k}}\}$, $\mathcal{B}= \mathcal{K}(t)\backslash\{\text{\MU{K}}\}$}
\State{\bfseries Step 2. }
\parWhile{$\mathcal{B} \neq \{\}$}
\State{\bfseries Step 2.1 Propose Phase} \parState{Following the sequence of $\boldsymbol{p}$, each \MU{i} in $\mathcal{B}$ proposes to the MUs in $\mathcal{A}$ according to the row $\boldsymbol{P}_i$}
\State{\bfseries Step 2.2  Accept or Reject Phase} \parState{Each \MU{j} in $\mathcal{A}$ will accept \MU{i} if one of these conditions is satisfied: $R_{S,ki}\geq R_{L,k}$ or $R_{S,ji}\geq R_{S,js_j}$.}
\EndparWhile
\end{algorithmic}
\end{algorithm}
}
\subsubsection{Scheduling Mechanism}
\label{game-model}
\label{sec:optimization}
Since MUs are selfish and rational, we can use game theory to model their behavior while forming the D2D LAN. Using this model and Algorithm \ref{alg:estimation}, we design a scheduling mechanism for the BS which will provide incentive for MUs to participate in the D2D LAN and reduce their total energy consumption.

We formulate the interactions of MUs by a non-cooperative game denoted by the triplet $\mathcal{G}=(\mathcal{K}(t),\mathcal{A},u)$, where $\mathcal{K}_t$ is a finite set of $\Ksize$ MUs, indexed by $k$, ${\mathcal{A}=\mathcal{A}_1\times...\times \mathcal{A}_{\Ksize}}$ is the strategy space, and ${\mathcal{A}_k=\{C,D\}}$ is the strategy set for each \MU{k} where strategy $C$ corresponds to cooperating and relaying while strategy $D$ corresponds to defecting and not relaying.
Each vector $\boldsymbol{a} = (a_1, . . . , a_{\Ksize})\in \mathcal{A}$ is known as the strategy profile and $\boldsymbol{u} = (u_1, . . . , u_{\Ksize})$ where ${u_k:\mathcal{A}_k \rightarrow \mathbb{R}}$ is the payoff function for \MU{k}.  The payoff of each MU is taken to be the negative of its total energy consumption.
We assume that the end of the real-time content is not previously known for anyone. Payoffs are calculated at each time slot and depend on the strategy profile played by MUs at that time slot. Here we consider two strategy profiles, namely, all MUs cooperate (All-C) and all MUs defect (All-D) and calculate MUs' payoffs under these profiles. Here, \MU{k}'s payoff is denoted by $u_k(a_k,a_{-k})$, where $a_k$ is the strategy chosen by \MU{k} and $a_{-k}$ represents strategies of the remaining MUs. Therefore, the payoff of \MU{k} at time slot $t$ is:
\begin{align}
u_k(D,D...,D) =-E_{k},\;
u_k(C,...,C) =-E'_{k},
\label{Ucoop}
\end{align}
where the $E_{k}$ and $E_{k}'$ are calculated from (\ref{Emulti}) and (\ref{Etotal}).
If \MU{k} defects and other MUs cooperate, \MU{k} will only receive data and its payoff will be:
\begin{align}
u_k(D,C...,C)= -\left(\rho_kP_{Rx}+\hspace{-4mm}\sum_{m\in{\mathcal{K}(t)\backslash\{k\}}}\hspace{-4mm} \rho_m \scr{P}_{Rx}\right)T.
\label{Udefcoop}
\end{align}
Comparing (\ref{Ucoop}) and (\ref{Udefcoop}), we see that
each MU prefers to be in the D2D LAN but always as a sink. To solve this game, we must find the Nash equilibrium (NE). An NE is defined as a strategy profile of the game in which no single MU can increase its payoff by unilateral deviation. For this game, the NE will be All-D because the payoff of cooperating is lower than defecting, no matter how other MUs choose their strategy.
However, when this ``stage game'' is repeated in every time slot~(stage) of the content, since the end of the content is unknown, there is a possibility that the outcome of this new ``infinitely repeated game'' would be different. In Subsection \ref{sec:SPM}, we discuss how the All-C profile is enforced as the NE of the repeated game.
The scheduling mechanism reduces the total energy consumption of MUs and guarantees that each MU consumes less energy if it joins the D2D LAN (an incentive). To achieve this, the BS, as the mediator, solves an optimization problem to find the seed time of each MU. Next, after defining a needed constraint, we discuss the optimization problem in detail.

\begin{definition}{}
\emph{Individual Rationality Constraint (IRC)}~\cite{shoham2008multiagent}: the mechanism $\chi_{BS}$ for MCRCD protocol is said to be \emph{individually
rational} if for every \MU{k} in $\mathcal{K}_t$ we have $\nonumber u_k(C,...,C)\geq u_k(D,...,D)\text{ or }
E_{k}' \leq E_k $.
\end{definition}
Constraints on individual rationality lead to an upper bound on \MU{k}'s seed time $\rho_k T$. The goal of the scheduling mechanism $\chi_{BS}$ is to minimize the total power consumption of the MUs and to motivate them to participate in the D2D LAN. Therefore, the problem of designing the scheduling mechanism can be represented as an optimization problem subject to the IRC and a finite set of other linear constraints:
\begin{align}
&\underset{ \rho_k :\forall k\in \mathcal{K}(t)}{\text{min}}\sum_{k\in \mathcal{K}(t)} E'_k(t) \label{opt_problem}\\
\nonumber&\text{s.t. }\forall k\in \mathcal{K}(t),\; \rho_k \geq 0,\; \sum_{k\in \mathcal{K}(t)} \rho_k=1,   E_{k}' \leq E_k.
\end{align}
The first and the second constraints make sure that every MU is selected as the seed at some point during each time~slot.

The last constraint ensures that IRC is satisfied for all MUs. $\rho_k T$ is the time duration of being seed for $\text{MU}_k$. In this formulation, we have assumed that during each time slot, channel conditions stay the same and that the network topology does not change. Moreover, the protocol does not require a-priori knowledge of mobility of the MUs and makes decisions based on the current $\mathcal{K}$ and the channel states.

\subsection{Self-organizing D2D LAN Formation}
\label{sec:SOD2D}
The MUs will autonomously engage in
D2D LAN formation. Once the BS has announced the seed \MU{k}, proposal order $\boldsymbol{p}$ and the seed times $\rho_iT$s, MUs carry out the graph formation autonomously. That is, in the sub-slot where \MU{k} is the seed, \MU{p_1} proposes to \MU{k}; \MU{k} checks the condition $R_{S,k\hspace{0.1em}p_1}\geq R_{L,k}$ and if it's not satisfied, rejects \MU{p_1} which must then act alone. However, if it is satisfied \MU{i} is added to $\mathcal{A}$ and \MU{p_2} gets a chance to propose. The following MUs, namely, $\text{\MU{p_2}},\dotsc,\text{\MU{p_K}}$ have progressively more options to propose to and they do so according to their corresponding preference lists which form the rows of~$\boldsymbol{P}$.
\subsection{Self-punishment Mechanism}
\label{sec:SPM}

In the next phase, the MUs will adopt a self-punishment mechanism and punish any non-abiding MU that refrains from relaying.
Using backward induction, we can prove that All-D is an NE of the repeated game whose stage game is $\mathcal{G}$\cite{shoham2008multiagent}.

In this subsection, we provide a mechanism that not only ensures that All-C is also an NE but also forces the game into either All-C or All-D profiles. This mechanism guarantees cooperative behavior throughout a time slot.
Let $p_{k}^{(t+1)}$ be the probability that \MU{k} assigns to the existence of the next time slot ($t+1$).
$p_{k}^{(t+1)}$ represents the uncertainty that MUs have about the termination of the content in the next time slot, as well as their uncertainty about other MUs' continued presence in the D2D LAN.
We assume that at any given time slot $t_0$, $p_{k}^{(t_0+1)}=p_{k}^{(t_0+2)} =...= p_{k}^{(t_0+N)},N \in \mathbb{N}$. Consequently, the interactions of MUs from any time slot onwards can be modeled by an infinitely repeated game.
In other words, we assume that there are an unknown number of time slots of content and that in each, the stage game $\mathcal{G}$ is being played; this repeated game may terminate at the end of time slot $t$ with probability $1-p_{k}^{(t+1)}$. Therefore, if all MUs cooperate until an unknown end, the total discounted payoff of each \MU{k} is calculated~as:
\begin{equation}
U_k(C,...,C)=\sum_{t=1}^{\infty}u_k^t(C,...,C)= \sum_{t=1}^{\infty}(p_k)^{t-1}E'_k(t),
\label{Urepeat}
\end{equation}
where, $E'_k$ is calculated according to (\ref{Etotal}) and $p_k$ is the probability that  \MU{k} assigns to the existence of the next time slot at the beginning of the game.
According to the folk theorem, we can use a grim-trigger punishment to guarantee that the All-C is the NE of this infinitely repeated game~\cite{shoham2008multiagent}.
\begin{definition}
\emph{Grim-trigger punishment $(P^{GT})$}: At time slot $t+1$, \MU{k} will defect unless \emph{every} MU (including itself) has cooperated during time slot $t$.
\end{definition}
Ideally, all MUs relay and therefore the game remains in the All-C equilibrium. If, however, one of MUs does not relay, there will be no longer any D2D LAN for the remaining time slots of the content.~
If the D2D LAN is to be continued in the time slot ${t+1}$, $p_{k}^{(t+1)}$ must be equal to or exceed a threshold value, which we call the critical expectation value (CEV). Specifically, if this condition holds, for every MU, choosing to cooperate is a ``best response'' as discussed next.
\begin{lemma} If at time slot~$t$, \MU{k}'s expectation about the existence of the next time slot is equal to or higher than
the critical expectation value $p_{k}^{*}$, the Cooperation strategy is the best response of \MU{k} during time slot $t$ in the MCRCD protocol.
The \emph{CEV} is calculated~using:
\begin{align}
p_k^{*}=\frac{E'_k-(P_{Rx}-\scr{P}_{Rx})\rho_kT-\scr{P}_{Rx}T}{E_k-(P_{Rx}-\scr{P}_{Rx})\rho_kT+\scr{P}_{Rx}T}.
\label{CVS}
\end{align}
\end{lemma}
{\small
\begin{proof}
For ``cooperation'' to be the best response of MUs to each other in the context of the repeated game described in \ref{sec:SPM}, \MU{k} should have a strong-enough belief in the existence of the next time slot $t$. If MUs use $P^{GT}$ to enforce All-C we have: $E_k(D,P^{GT})~\geq~E_{k}(C,C)$, consequently,

\begin{align*}
&P_{Rx}\rho_k T+\scr{P}_{Rx}(1-\rho_k)T+\sum_{n=1}^{\infty}(p_k)^n E_k\geq
 \sum_{n=0}^{\infty}(p_k)^n E'_k,\\
&(P_{Rx}-\scr{P}_{Rx})\rho_kT+\scr{P}_{Rx}T+\frac{p_k}{1-p_k} E_k \geq
\frac{1}{1-p_k}E'_k,\\
&p_k \geq \frac{E'_k-(P_{Rx}-\scr{P}_{Rx})\rho_kT-\scr{P}_{Rx}T}{E_k-(P_{Rx}-\scr{P}_{Rx})\rho_kT+\scr{P}_{Rx}T}.
\end{align*}
Note that, at any given time slot $t_0$, $p_{k}^{(t_0+1)}=p_{k}^{(t_0+2)} =...= p_{k}^{(t_0+N)},N \in \mathbb{N}$. In other words, \MU{k} assumes that the current state of the network does not change in the future time slots, and that in the upcoming time slots, $E_k$ and $E'_k$ will have the same values as they do in the current time slot.
\end{proof}
}
\section{Simulation Results}
\label{Sec:Simulation}
For our simulations, we consider a set of $K$ MUs randomly deployed in a
$400$~m $\times$ $400$~m area with the BS located at the center of the area.
The total number of RBs is 25. The BS transmission power is 5 W which is equally divided among RBs. The maximum number of allowed hops is set to $H = 4$ hops.
To mitigate the effect of interference, we assume that the BS allocates resources in a manner that the maximum power of each MU on the SR link equals $125$ mW. Meanwhile, the maximum interference is restricted to $0.01\%$ of the received power. The energy consumed per unit time can be considered to be almost constant for various transmission bit rates using adaptive rate control~\cite{al2014optimal}. Following these studies, we set these parameters to the following values: $P_{R,L}(R) = 1.8$ W, $P_{R,S}(R) = 0.925$ W and $P_{T,S}(R)=1.425$ W. The thermal noise $\sigma^2$ is considered to be $10^{-13}$ mW.
In all simulations, we repeat the simulation to get confidence interval levels of approximately $95\%$ with the variance equal to $1.5$. Throughout this section, we compare three schemes: the optimal scenario, the MCRCD protocol and the multicast scenario illustrated in Fig. \ref{system model}.
\begin{figure*}[t]
\centering
\subcaptionbox{\label{MUs_number:throughput}}{\includegraphics[height=1.35in, width=0.32\textwidth]{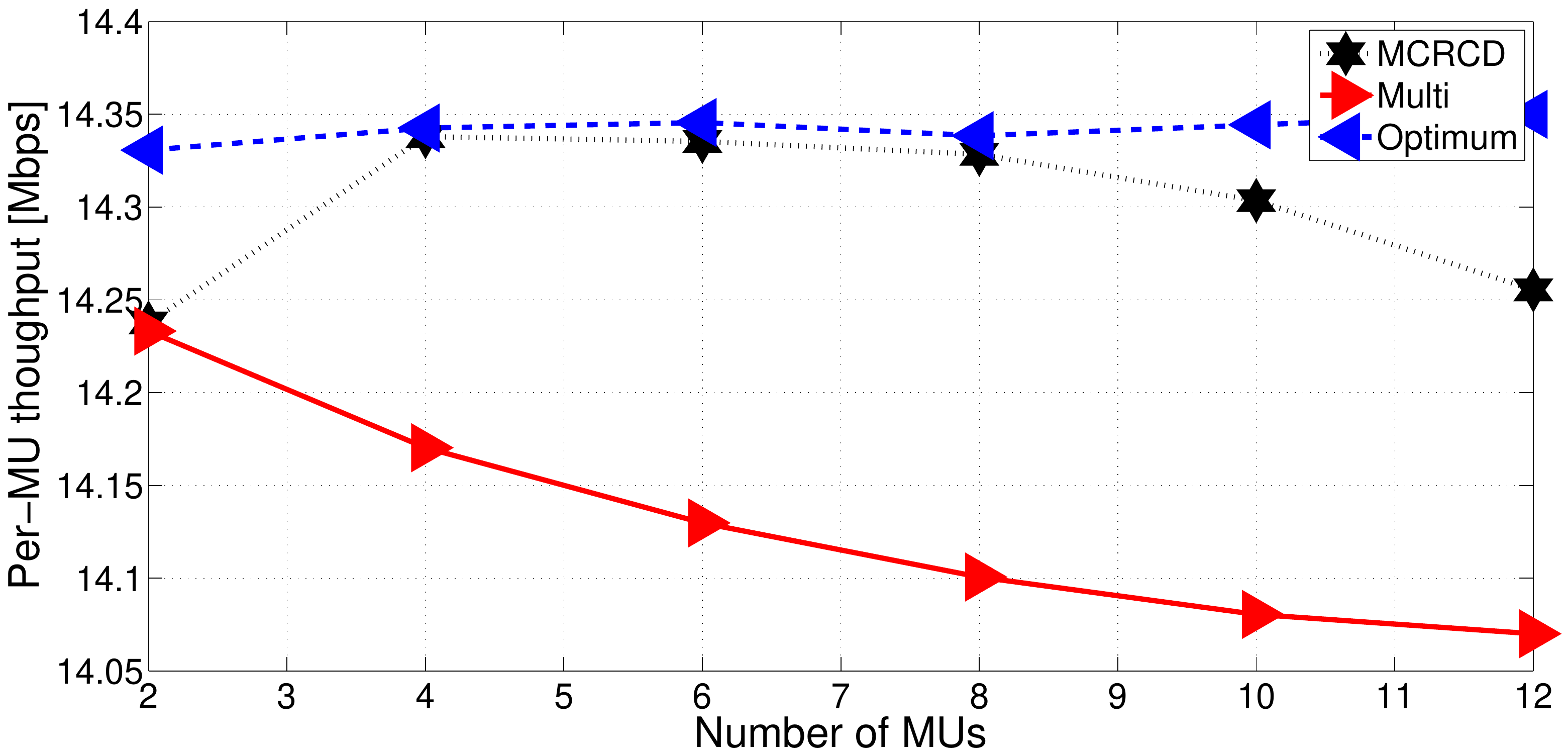}‎}
\subcaptionbox{\label{MUs_number:energy_efficiency_avg}}{\includegraphics[height=1.35in, width=0.32\textwidth]{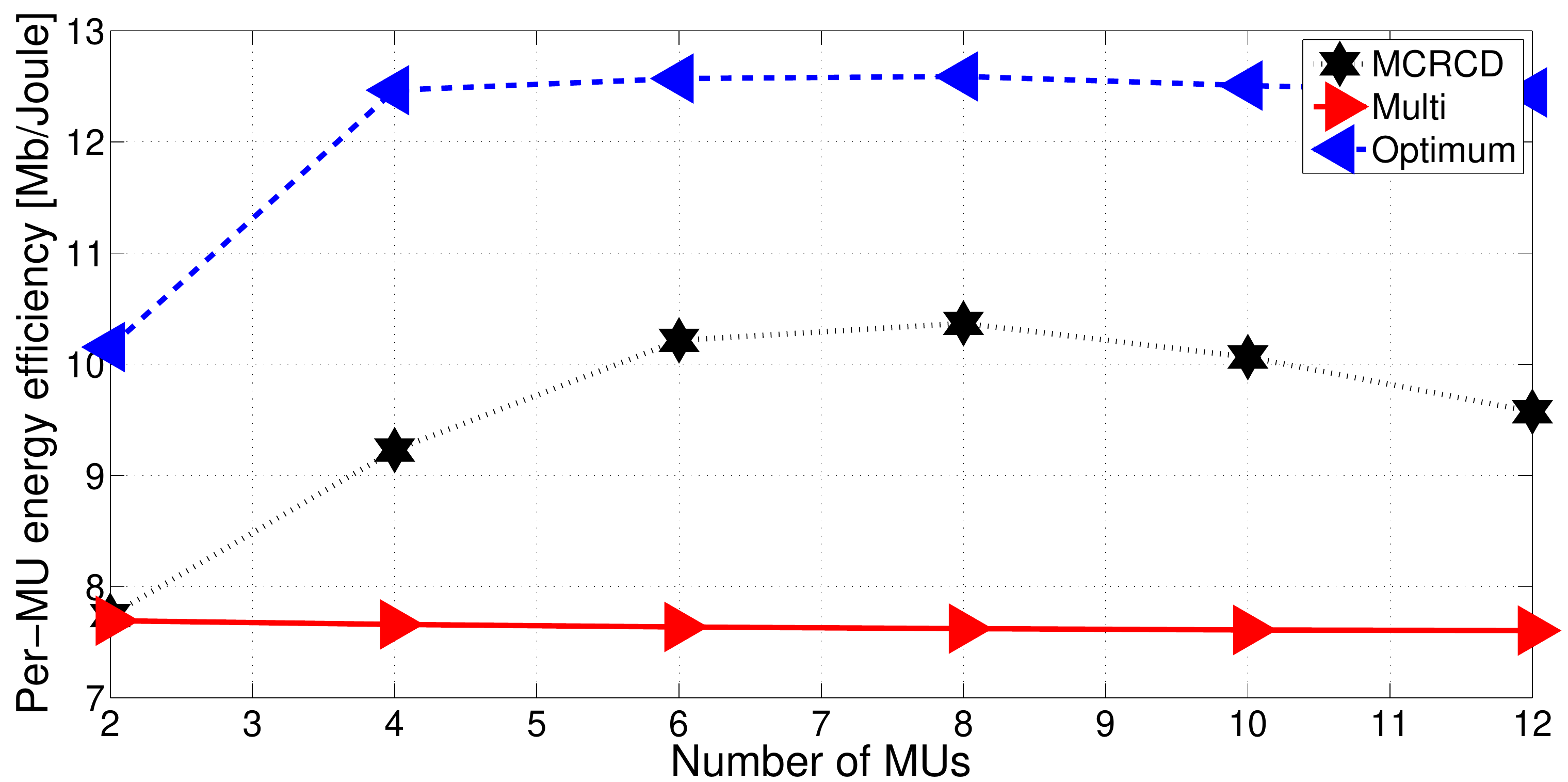}‎
}
\subcaptionbox{\label{CEV}}{\includegraphics[height=1.35in, width=0.32\textwidth]{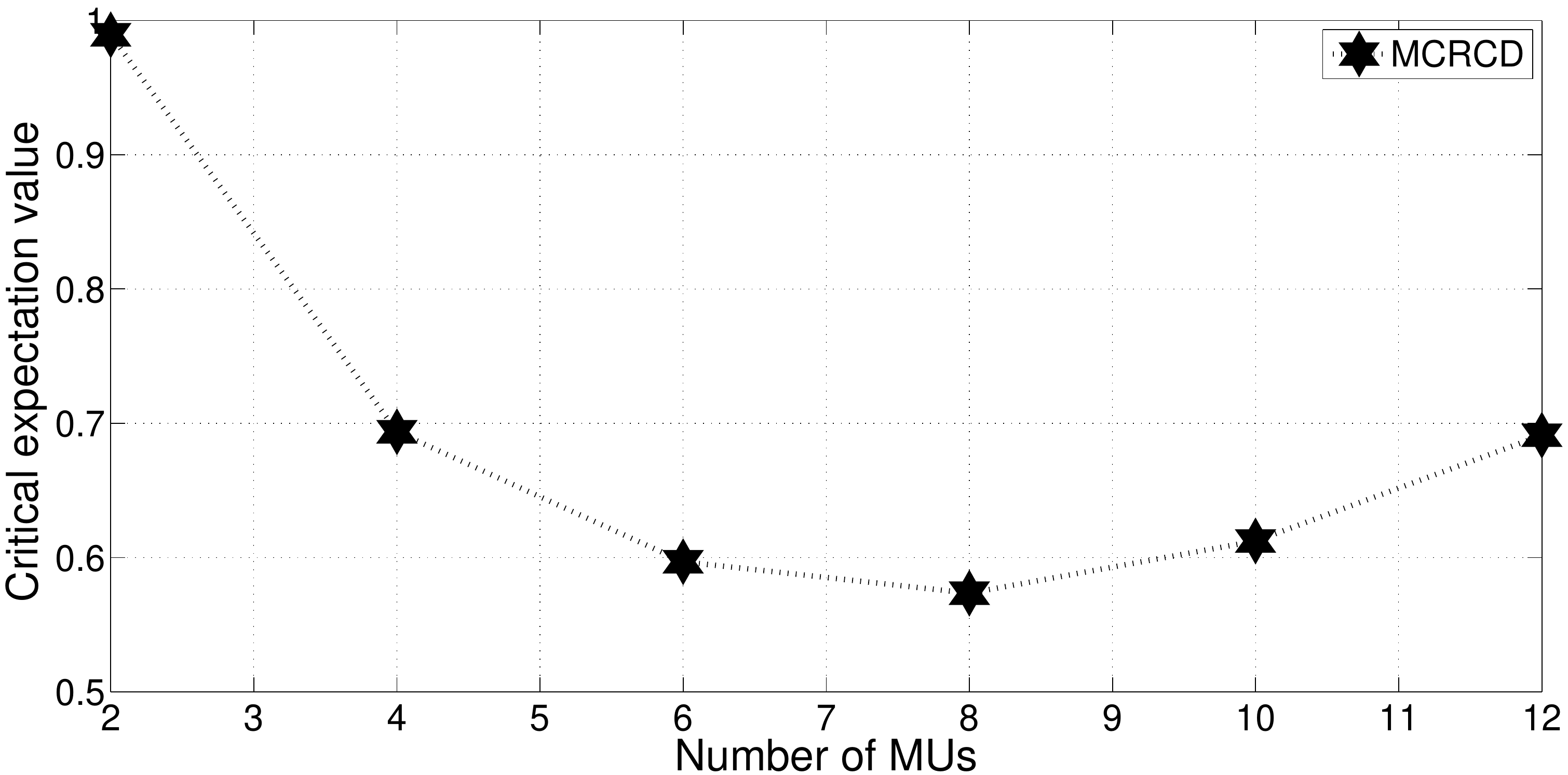}‎}
\caption{The effect of the number of MUs‎ on average throughput (\ref{MUs_number:throughput}), energy efficiency (\ref{MUs_number:energy_efficiency_avg}), and  critical expectation value(\ref{CEV}).}
\label{MUs_number:1}
  \vspace{-1.5em}
\end{figure*}

In Fig. \ref{MUs_number:1}, we evaluate the impact of the number of MUs co-located in the given area on the performance of the MCRCD.
Fig. \ref{MUs_number:throughput} shows that in the multicast scenario, average per-MU throughput is decreasing in
$\Ksize$; this is because the extra MUs in $\mathcal{K}_t$ can at best maintain the minimum LR bit rate, while it's more likely that they would decrease it. The average per-MU throughput of the optimal scenario does not change due to the fact that always the optimal seeds are selected.
Under the MCRCD protocol, average per-MU throughput decreases in
$\Ksize$, but it closely follows the optimal scenario. The reason is that all MUs are being scheduled and the ones with unfavorable channel states affect the average throughput less severely.
In terms of per-MU energy efficiency, i.e., the amount of data received per unit energy consumed [bits/Joule], Fig.~\ref{MUs_number:energy_efficiency_avg}~shows that MUs obtain higher energy efficiencies under the cooperative, optimal and MCRCD scenarios compared to the non-cooperative multicast scenario. This is due to the fact that the cooperative scheme not only increases throughput but also decreases power consumption per MU.
In the MCRCD protocol, MUs are taken to be selfish, therefore the energy efficiency per MU is smaller than the optimal scenario.
The reason for the trend shift in Fig.~\ref{MUs_number:energy_efficiency_avg} when $K$ reaches and goes beyond 10 MUs, is that as $K$ increases, the IRC
becomes harder to satisfy for every MU in the D2D LAN. Consequently, the optimization problem (\ref{opt_problem}) becomes infeasible, i.e., a single inclusive D2D LAN will be no longer possible. At this point, either some MUs can form a new D2D LAN, or the D2D LAN breaks up and the MUs continue receiving under the multicast scenario. In this paper, we have assumed the latter.
In Fig. \ref{CEV}, the critical expectation value (CEV) is plotted versus the number of MUs. By increasing $\Ksize$, the CEV decreases. This shows how an increase in the number of MUs, relaxes their concerns about the continuation of the game, i.e., the D2D LAN and make them more \emph{patient}. From Fig. \ref{MUs_number:energy_efficiency_avg}, the average per-MU energy efficiency is increasing, therefore MUs prefer playing All-C profile more when $\Ksize$ increases.
\section{Conclusion}
\label{Sec:Conclusion}
In this paper, we have proposed a novel energy-aware multi-hop protocol for optimizing the formation of D2D LANs in emerging cellular networks. In the studied model, we have focused on the selfish behavior of MUs. In our model, the BS is a mediator that assists selfish MUs in cooperating and forming a D2D LAN. We have proposed a game-theoretic approach to model D2D LAN formation and cooperation in data relaying. We have introduced a graph estimation algorithm to simulate the interactions of MUs and estimate the resulting graph. To provide incentive for participation in the D2D LAN, we have developed a mechanism that reduces the total energy consumption of MUs. Finally, to enforce cooperative behavior (relaying) in all time slots, a self-punishment mechanism was proposed. Simulation results have shown the various merits of the proposed approach along with the corresponding gains.

\bibliographystyle{IEEEtran}
\bibliography{D2DLAN}

\begin{thebibliography}{1}
\providecommand{\url}[1]{#1}
\csname url@samestyle\endcsname
\providecommand{\newblock}{\relax}
\providecommand{\bibinfo}[2]{#2}
\providecommand{\BIBentrySTDinterwordspacing}{\spaceskip=0pt\relax}
\providecommand{\BIBentryALTinterwordstretchfactor}{4}
\providecommand{\BIBentryALTinterwordspacing}{\spaceskip=\fontdimen2\font plus
\BIBentryALTinterwordstretchfactor\fontdimen3\font minus
  \fontdimen4\font\relax}
\providecommand{\BIBforeignlanguage}[2]{{%
\expandafter\ifx\csname l@#1\endcsname\relax
\typeout{** WARNING: IEEEtran.bst: No hyphenation pattern has been}%
\typeout{** loaded for the language `#1'. Using the pattern for}%
\typeout{** the default language instead.}%
\else
\language=\csname l@#1\endcsname
\fi
#2}}
\providecommand{\BIBdecl}{\relax}
\BIBdecl

\bibitem{Asadi2014}
A.~Asadi, Q.~Wang, and V.~Mancuso, ``A survey on device-to-device communication
  in cellular networks,'' \emph{Communications Surveys Tutorials, IEEE},
  vol.~16, no.~4, pp. 1801--1819, Fourthquarter 2014.

\bibitem{alexact}
L.~Al-Kanj and Z.~Dawy, ``{Exact and Heuristic Solutions for Energy-Aware
  Multihop Cooperation Over Wireless Networks},'' \emph{IEEE Trans. on
  Vehicular Technology}, vol.~64, no.~7, pp. 2952--2971, July 2015.

\bibitem{al2014optimal}
L.~Al-Kanj, H.~Poor, and Z.~Dawy, ``{Optimal Cellular Offloading via
  Device-to-Device Communication Networks With Fairness Constraints},''
  \emph{IEEE Wireless Communications}, vol.~13, no.~8, pp. 4628--4643, Aug
  2014.

\bibitem{al2008optimal}
Y.~Al-Chikhani and S.~Sami, ``{On optimal video distribution over
  infrastructure controlled P2P networks},'' Ph.D. dissertation, 2008.

\bibitem{popova2008cooperative}
L.~Popova, T.~Herpel, W.~Gerstacker, and W.~Koch, ``Cooperative
  mobile-to-mobile file dissemination in cellular networks within a unified
  radio interface,'' \emph{Computer Networks (Elsevier)}, vol.~52, no.~6, pp.
  1153--1165, 2008.

\bibitem{goldsmith2005wireless}
A.~Goldsmith, \emph{Wireless communications}.\hskip 1em plus 0.5em minus
  0.4em\relax Cambridge university press, 2005.

\bibitem{albano2011context}
M.~Albano, M.~Alam, A.~Radwan, and J.~Rodriguez, ``Context aware node discovery
  for facilitating short-range cooperation,'' in \emph{Proc. of 26th Wireless
  World Research Forum (WWRF26)}, Doha, Qatar, Apr. 2011.

\bibitem{shoham2008multiagent}
Y.~Shoham and K.~Leyton-Brown, \emph{Multiagent systems: Algorithmic,
  game-theoretic, and logical foundations}.\hskip 1em plus 0.5em minus
  0.4em\relax Cambridge University Press, 2008.

\bibitem{kim2014full}
S.~Kim and W.~Stark, ``{Full duplex device to device communication in cellular
  networks},'' in \emph{Proc. of IEEE Int. conf. Computing, Networking and
  Communications (ICNC)}, Honolulu, Hawaii, Feb. 2014, pp. 721--725.

\end{thebibliography}
\appendices
\balance
\end{document}